\documentclass[aps,prl,preprint,footinbib]{revtex4-1}

\usepackage{graphicx}
\usepackage{hyperref}
\usepackage{amsmath,amssymb}

\begin{document}


\title{Unified theoretical description of thermal and nonthermal laser-induced ultrafast structural changes in solids}

\renewcommand{\figurename}{FIG.}
\renewcommand{\tablename}{TAB.}



\author{Bernd Bauerhenne}
\email{bauerhenne@uni-kassel.de}
\author{Martin E. Garcia}
\affiliation{Institute of Physics and Center for Interdisciplinary Nanostructure Science and Technology (CINSaT), University of Kassel, Heinrich-Plett-Strasse 40, 34132 Kassel, Germany}


\date{\today}

\begin{abstract}
The ultrafast dynamics of ions in solids following intense femtosecond laser excitation is governed by two fundamentally distinct yet interplaying effects.
On one hand, the significant generation of hot electron-hole pairs by the light pulse alters the strength and nature of interatomic bonding, resulting in nonthermal ionic motion. 
On the other hand, incoherent electron-ion collisions drives an equilibration of the electrons and the ions, which reach a common temperature on a picosecond timescale. 
This letter introduces, for the first time, a unified theory that comprehensively accounts for both phenomena. 
Our approach is versatile, applicable to both ab-initio and large-scale molecular dynamics simulations, and leads to a generalization of the two-temperature model-molecular-dynamics equations of motion.
To illustrate the effectiveness of our method, we apply it to describe the laser excitation of silicon thin films. Our simulations reproduce the time evolution of the Bragg peaks in excellent agreement with the experiments.
\end{abstract}


\maketitle
Upon exposure to an intense femtosecond (fs) laser pulse, a material undergoes a complex interplay of competing ultrafast processes.
 Given the strong coupling of the laser field with electrons and its comparatively weak interaction with ions,
 it is widely accepted \cite{Stampfli1990,Stampfli1994} that a short-lived non-equilibrium state is created, which consists of hot electron hole pairs and relatively cold ions.
 
Incoherent electron-phonon collisions induce an energy transfer from the electrons to the ions, leading to an equilibration of the electronic- $T_\text{e}$ and ionic temperature $T_\text{i}$ on a picosecond timescale $\tau_\text{ep}$, effectively relaxing the non-equilibrium state.
 The thermal impact of electron-phonon coupling (EPC) on laser-induced structural phenomena has been simulated employing the two-temperature-model molecular-dynamics simulation (TTM-MD) method, relying on empirical interatomic potentials 
 \cite{Hakkinen1993,Ivanov2003}:
\begin{eqnarray}
    C_\text{e}(T_\text{e}) \frac{dT_\text{e}}{dt} &=&  G_\text{ep}(T_\text{e})\, (T_\text{e} - T_\text{i}) + \frac{dE_{\text{L}_\text{abs}}}{dt}, \label{equ:e_TTM_MD}\\
    M_k\, \frac{d^2\mathbf{r}_k}{dt^2} &=& -\nabla_{\mathbf{r}_k} U + \xi \, M_k\, \mathbf{v}_k, 
    \label{equ:ionic_TTM_MD}
\end{eqnarray}
where $C_\text{e}(T_\text{e})$ refers to the electronic heat capacity, $G_\text{ep}(T_\text{e})$ to the electron-phonon coupling constant. $ dE_{\text{L}_\text{abs}} / dt$ describes the energy flow absorbed from the laser. $-\nabla_{\mathbf{r}_k} U$ is the force acting on atom $k$ with mass $M_k$, coordinate $\mathbf{r}_k$ and velocity $\mathbf{v}_k$ evaluated from the interatomic potential or potential energy surface (PES) $U(\mathbf{r}_1,\ldots)$. The "friction" constant $\xi$ is derived from energy conservation \cite{Ivanov2003}. 
For the sake of simplicity, Eq. \eqref{equ:e_TTM_MD} and \eqref{equ:ionic_TTM_MD} assume homogeneous spatial temperature profiles. 
The current TTM-MD methodology overlooks a crucial aspect—the impact of hot electrons on interatomic bonding during the laser-induced non-equilibrium state. 
Typically, this non-thermal influence is modeled by calculating the potential energy surface (PES) for electrons at elevated temperatures. This "laser-excited" PES  essentially governs the ionic motion immediately after the laser-excitation in many solids.
State-of-the-art ab initio approaches yield dramatic changes of the PES upon laser excitation, so that ions move initially non-thermally. 
Ultrafast nonthermal motions lead to structural phenomena that cannot occur in thermodynamic equilibrium as ultrafast  phase transitions \cite{Cavalleri2001,Sciaini2009,Buzzi2018,Yijing2022}, thermal phonon squeezing \cite{Johnson2009,Zijlstra2013Jan} and coherent phonons \cite{Cheng1991,Hase2003,Curcio2022}.
The physics of the influence of the PES at high $T_\text{e}$ has been already described microscopically by many groups using finite temperature density functional theory (DFT) \cite{Recoules2006,Zijlstra2013Aug}.
The microscopic description of EPC has also been successfully achieved on the basis of the ab-initio determination of the Eliashberg function \cite{Lin2008,Waldecker2016,Krylow2020} and other approaches \cite{Sadasivam2017,Nikita2023}.
The processes driven by the EPC and by the laser-induced alterations of the PES are competing and should therefore be treated on the same microscopic theoretical level.
Moreover, recent experiments \cite{Munoz2023} underscore the necessity of considering both effects to elucidate laser-induced ultrafast lattice disordering. 
Despite some efforts to consider both influences \cite{Shokeen2010,Darkins2018,Lian2016}, no unified first principles theory was developed so far.
In this letter, we present a first-principles derivation of the equations of motions of ions including the competing effects of EPC and excited PES.
We  generalize, both the TTM-MD \cite{Ivanov2003} method, enabling it to account for the influence of the excited PES, and the method of  ab-initio simulations based on  $T_\text{e}$-dependent PES \cite{Silvestrelli1996,Recoules2006,Zijlstra2013Aug}, allowing it to consider  the influence of EPC as well.

All ab initio approaches determining the laser-excited PES $\Phi$ are based on the Mermin Free energy \cite{Mermin1965,Silvestrelli1996} of the electrons at elevated $T_\text{e}$
\begin{equation}
    \Phi(T_\text{e}, \mathbf{r}_1,\ldots) = U(\mathbf{r}_1,\ldots) - T_\text{e} \, S_\text{e}(T_\text{e}, \mathbf{r}_1,\ldots),
\end{equation}
where $S_\text{e}$ is the electronic entropy and $U$ is the PES at $T_\text{e}=0$.
Here, the electrons are treated in the canonical ensemble.
In the usual TTM-MD approaches, the ions are treated in the microcanonical ensemble, which ensures energy conservation, although there is an energy transfer between electrons and ions due to the EPC.
In most ab initio approaches one assumes a constant $T_\text{e}$ which allows MD simulations in the microcanonical ensemble for the ions using $\Phi$.
However, the total energy is not conserved, because the complete system of ions and electrons is treated in a mixed ensemble (microcanoncal and canonical).
As soon as the EPC is considered, $T_\text{e}$ changes with time and the ions cannot be treated any more in the microcanonical ensemble, since the sum of ionic potential energy $\Phi$ and kinetic energy $E_\text{kin}$ is not conserved.
Therefore, in order to be able to treat $\Phi(T_\text{e},\mathbf{r}_1,\ldots)$ and EPC in a unified way, one has to transform the electrons to the microcanonical ensemble.
In the following, we explain how to implement this idea.

We consider a solid consisting of $N_\text{at}$ atoms of the same type with mass $M$ excited by a fs laser pulse, so that the electrons reach a uniform $T_\text{e}$.
We include all coordinates $\mathbf{r}_i$ in the vector $\vec{R} \in \mathbb{R}^{3\, N_\text{at}}$ and all velocities $\mathbf{v}_i$  in $\vec{V} \in \mathbb{R}^{3\, N_\text{at}}$.
We denote the PES of the ions with electrons at $T_\text{e}$ by $\Phi(\vec{R},T_\text{e})$, so that the PES related force on atom $i$ is given by $- \nabla_{\mathbf{r}_i} \Phi$.
The vector $-\nabla_{\vec{R}} \Phi \in \mathbb{R}^{3\, N_\text{at}}$ contains all atomic forces.
$S_\text{e}$ and  $C_\text{e}$ can be directly calculated from $\Phi$ by
\begin{eqnarray}
	S_\text{e} &=& -\frac{\partial \Phi}{\partial T_\text{e}}, \qquad
	C_\text{e} = -T_\text{e} \, \frac{\partial^2 \Phi}{\partial T_\text{e}^2}.
	\label{equ:C_e_S_e}
\end{eqnarray}
The ions do not exhibit a well defined temperature immediately after the laser exictation, since their motion is mainly determined by $\Phi$.
Nevertheless, using the velocities of the ions and the equipartition theorem, one can assign an average "ionic temperature" $T_\text{i}$ to the ions.
Or in general, one can assign individual ionic temperatures for sets of phonon modes \cite{Waldecker2016}.
By diagonalizing the dynamical matrix, one obtains the polarization vectors $\vec{e}^{\;(1)},\ldots, \vec{e}^{\;(3\,N_\text{at})} \in \mathbb{R}^{3\,N_\text{at}}$ of all phonon modes.
The polarization vectors are orthonormal and form a complete basis set of the $\mathbb{R}^{3\, N_\text{at}}$.
We divide the phonon modes into $N_\mathcal{M}$ disjoint subsets $\mathcal{M}_k$.
We denote by $|\mathcal{M}_k|$ the number of corresponding phonon modes in set $\mathcal{M}_k$.
For each set $\mathcal{M}_k$, we define the corresponding projection operator $\mathbf{P}_{\mathcal{M}_k} \in \mathbb{R}^{3\,N_\text{at} \times 3\,N_\text{at}}$ by
\begin{equation}
	\mathbf{P}_{\mathcal{M}_k} = \sum_{j \in \mathcal{M}_k} \vec{e}^{\;(j)} \cdot \left(\vec{e}^{\;(j)}\right)^\text{t}.
	\label{equ:P_M_k}
\end{equation}
This operator projects the atomic velocities $\vec{V}$ onto the directions of the phonon modes of set $\mathcal{M}_k$, obeys $\mathbf{P}_{\mathcal{M}_k}^t = \mathbf{P}_{\mathcal{M}_k} = \mathbf{P}_{\mathcal{M}_k}^2$, and is used to define an individual ionic temperature for a given set $\mathcal{M}_k$ of phonon modes:
\begin{equation}
	T_{\text{i}_{\mathcal{M}_k}} = \frac{2\, E_{\text{kin}_{\mathcal{M}_k}}}{|\mathcal{M}_k| \, k_\text{B}} = \frac{M\, \vec{V}^{\;\text{t}} \cdot \mathbf{P}_{\mathcal{M}_k} \cdot \vec{V}}{|\mathcal{M}_k| \, k_\text{B}}.
	\label{equ:Ti_Mk}
\end{equation}
Here $k_\text{B}$ denotes the Boltzmann constant and $E_{\text{kin}_{\mathcal{M}_k}}$ the kinetic energy of the phonon modes of set $\mathcal{M}_k$.

Now we consider the total energy of the system as a function of time.
If $T_\text{e}$ is constant, it holds that
\begin{equation}
	\Phi\bigl.\bigr|_{T_\text{e}}+ E_\text{kin} = \text{const.},
	\label{equ:Energy_conservation_Phi_Te_const}
\end{equation}
where $E_\text{kin}$ is the total kinetic energy of the ions.
The ions are treated in the microcanoncial ensemble and no energy is needed to change $T_\text{e}$.
If $T_\text{e}$ is not constant, we must transform $\Phi$ to the internal energy of the electrons.
If the positions $\mathbf{r}_1,\ldots,\mathbf{r}_{N_\text{at}}$ of the ions are kept constant, we can easily do this transformation by
\begin{equation}
	E\bigl.\bigr|_{\mathbf{r}_1,\ldots,\mathbf{r}_{N_\text{at}}} = \underbrace{E_\text{kin}}_{=0} + \Phi + T_\text{e}\, S_\text{e} 
	\overset{\eqref{equ:C_e_S_e}}{=} \underbrace{E_\text{kin}}_{=0} + \Phi - T_\text{e}\,\frac{\partial \Phi}{\partial T_\text{e}}.
	\label{equ:U_ions_fixed}
\end{equation}
However, this transformation is incorrect for moving ions, since now the coordinates $\mathbf{r}_i$ are dynamical variables instead of parameters.
Thus, it is impossible to decompose the internal energy of the ions in an algebraic way from the Helmholtz free energy of the electrons.
$\Phi$ is the potential part of the internal energy of the ions and, at the same time, the Helmholtz free energy of the electrons.
Now only the electronic contribution must be transformed to the internal energy.
Remarkably, if one considers the time derivative of Eq \eqref{equ:U_ions_fixed}, one can disentangle the ionic and electronic contributions and can generate a transformation only affecting the electrons:
\begin{equation}
	\frac{dE}{dt} = \frac{E_\text{kin}}{dt} + \frac{d\Phi}{dt} -  \frac{d}{dt} \underbrace{\left( T_\text{e} \, \frac{\partial \Phi}{\partial T_\text{e}} \right)}_{\text{only e}^-}.
\end{equation}
The second term must only take the electronic degree of freedom $T_\text{e}$ into account, therefore we set
\begin{eqnarray}
    \frac{d}{dt} \underbrace{\left( T_\text{e} \, \frac{\partial \Phi}{\partial T_\text{e}} \right)}_{\text{only e}^-} &=& \frac{dT_\text{e}}{dt} \, \frac{\partial \Phi}{\partial T_\text{e}} + T_\text{e}\, \underbrace{\sum \limits_{j=1}^{N_\text{at}} \left(\nabla_{\mathbf{r}_j}\frac{\partial \Phi}{\partial T_\text{e}}\right)^\text{t} \cdot \frac{d\mathbf{r}_j}{dt}}_{\overset{!}{=}0} \nonumber \\
    &&+ T_\text{e}\,  \frac{\partial^2 \Phi}{\partial T_\text{e}^2} \,\frac{dT_\text{e}}{dt},
\end{eqnarray}
since the second term on the right hand side is related to an ionic contribution.
We obtain
\begin{eqnarray}
	\frac{dE}{dt}&=& 
 \frac{E_\text{kin}}{dt} + \frac{d\Phi}{dt} - \frac{dT_\text{e}}{dt} \, \frac{\partial \Phi}{\partial T_\text{e}} - T_\text{e}\,  \frac{\partial^2 \Phi}{\partial T_\text{e}^2} \,\frac{dT_\text{e}}{dt}.
	\label{equ:dE_dt_construction}
\end{eqnarray}
In order to determine the internal energy $E(t_\ell)$ of ions and electrons at time $t_\ell$ in a MD simulation starting at time $t_0$, we integrate Eq. \eqref{equ:dE_dt_construction} over time from $t_0$ to $t_\ell$.
We write $\Phi(t_\ell)$ instead of $\Phi\Bigl(T_\text{e}(t_\ell), \mathbf{r}_1(t_\ell), \ldots\Bigr)$ for readability.
Since the potential energy is only defined up to a constant, we can omit $E_\text{kin}(t_0)+\Phi(t_0)$ and get finally
\begin{equation}
	\boxed{E(t_\ell) = E_\text{kin}+ \Phi(t_\ell) - \int \limits_{t_0}^{t_\ell} dt \left(\frac{\partial \Phi}{\partial T_\text{e}} + T_\text{e}\,  \frac{\partial^2 \Phi}{\partial T_\text{e}^2} \right) \frac{dT_\text{e}}{dt}.}
	\label{equ:E_t1}
\end{equation}
Eq. \eqref{equ:E_t1} is the central equation of this letter, which allows to consider laser-induced changes of the PES and the EPC on the same ab initio level of theory.
Now we are able to formulate the energy conservation for moving ions and changing $T_\text{e}$.
Since the whole system is described in the microcanconical ensemble, we can consider the effect of the total energy absorbed from the laser up to time $t_\ell$ $E_{\text{L}_\text{abs}}(t_\ell)$ as
\begin{equation}
	E(t_\ell) = E_{\text{L}_\text{abs}}(t_\ell) + \text{const.}
	\label{equ:Energy_conservation_Phi}
\end{equation}
In order to derive the equations of motion, we calculate the time derivative of $E$ from Eq. \eqref{equ:E_t1}
\begin{eqnarray}
	\frac{dE}{dt} &=& M \sum_{j=1}^{N_\text{at}} \mathbf{v}^\text{t}_j \cdot \frac{d\mathbf{v}_j}{dt} +
 \sum_{j=1}^{N_\text{at}}  \mathbf{v}^\text{t}_j \cdot \nabla_{\mathbf{r}_j}\Phi - T_\text{e}\,  \frac{\partial^2 \Phi}{\partial T_\text{e}^2} \, \frac{dT_\text{e}}{dt} \nonumber \\
	&\overset{\eqref{equ:C_e_S_e}}{=}& M\, \vec{V}^{\;\text{t}} \cdot \frac{d\vec{V}}{dt} +\vec{V}^{\;\text{t}} \cdot \nabla_{\vec{R}} \Phi + C_\text{e} \, \frac{dT_\text{e}}{dt}.
	\label{equ:dE_dt}
\end{eqnarray}
The third term corresponds to the time derivative of the internal energy of the electrons
\begin{equation}
	\frac{dE_\text{e}}{dt} = C_\text{e} \, \frac{dT_\text{e}}{dt}.
	\label{equ:dEe_dt_Te}
\end{equation}
Since we neglect the local electronic heat flow by using a global $T_\text{i}$ and $T_\text{e}$, we consider only two processes that change the internal energy $E_\text{e}$ of the electrons:
The electrons can interchange energy with the ions due to EPC.
We denote the total energy that the electrons interchange with the ions up to time $t_\ell$ by $E_\text{ep}(t_\ell)$.
Since the electrons couple with different strengths to different phonon modes of the ions, we denote the different electron-phonon coupling constants by $G_{\text{ep}_{\mathcal{M}_k}}$ \cite{Waldecker2016}.
The constant $G_{\text{ep}_{\mathcal{M}_k}}$ indicates the size of the energy flow per phonon mode from the phonons of set $\mathcal{M}_k$ to the electrons depending on the temperature difference $T_\text{e}-T_{\text{i}_{\mathcal{M}_k}}$.
In addition, the electrons can absorb energy from a laser field. 
Hence, the time derivative of the internal energy of the electrons is given by
\begin{equation}
	\underbrace{C_\text{e} \, \frac{dT_\text{e}}{dt}}_{\overset{\eqref{equ:dEe_dt_Te}}{=}\frac{dE_\text{e}}{dt}} = -\sum_{k=1}^{N_\mathcal{M}}  |\mathcal{M}_k| \, G_{\text{ep}_{\mathcal{M}_k}} \left(T_\text{e} - T_{\text{i}_{\mathcal{M}_k}}\right) +  \frac{dE_{\text{L}_\text{abs}}}{dt}.
	\label{equ:Nat_Ce_dTe_dt}
\end{equation}
This is the differential equation to calculate the time change of $T_\text{e}$, which is also used in the two-temperature model 
\cite{Anisimov1974} and the TTM-MD \cite{Ivanov2003}, if only global ionic and electronic temperatures are considered.
The equations of motions for the ions is calculated from the time derivative of Eq. \eqref{equ:Energy_conservation_Phi}:
\begin{eqnarray}
	\frac{dE}{dt} &=& \frac{dE_{\text{L}_\text{abs}}}{dt}.
\end{eqnarray}
Inserting \eqref{equ:dE_dt} and using \eqref{equ:Nat_Ce_dTe_dt} for $C_\text{e} \, \frac{dT_\text{e}}{dt}$ we find
\begin{eqnarray}
 M\, \vec{V}^{\;\text{t}} \cdot \frac{d\vec{V}}{dt} + \vec{V}^{\;\text{t}} \cdot \nabla_{\vec{R}} \Phi && \nonumber \\
	 - \sum_{k=1}^{N_\mathcal{M}} \underbrace{\frac{M\, \vec{V}^{\;\text{t}} \cdot \mathbf{P}_{\mathcal{M}_k} \cdot \vec{V} }{2\, E_{\text{kin}_{\mathcal{M}_k}}}}_{=1} \, |\mathcal{M}_k| \, G_{\text{ep}_{\mathcal{M}_k}} \left(T_\text{e} - T_{\text{i}_{\mathcal{M}_k}}\right) &=& 0, \nonumber
\end{eqnarray}
The above equation must be valid for arbitrary velocities $\vec{V}$.
Therefore, it must hold that
\begin{equation}
	M\, \frac{d\vec{V}}{dt} = - \nabla_{\vec{R}} \Phi + \sum_{k=1}^{N_\mathcal{M}} \frac{|\mathcal{M}_k| \, G_{\text{ep}_{\mathcal{M}_k}} \left(T_\text{e} - T_{\text{i}_{\mathcal{M}_k}}\right)}{2\, E_{\text{kin}_{\mathcal{M}_k}}}\, M \, \mathbf{P}_{\mathcal{M}_k} \cdot \vec{V}.
	\label{equ:ionic_motion_global}
\end{equation}
The first term of the right hand side is the force coming from the PES at $T_\text{e}$ and the second term is the force coming from the electron-phonon coupling.

The set of equations \eqref{equ:Nat_Ce_dTe_dt} and \eqref{equ:ionic_motion_global} summarize the unified theory developed in this letter.
We analyze now two interesting limiting cases:
(1) If $T_\text{e}$ remains constant, $E(t_\ell)$ transforms to $E_\text{kin}+\Phi(t_\ell)$, since the integral term in Eq. \eqref{equ:E_t1} vanishes due to $dT_\text{e} / dt=0$.
Consequently, for constant $T_\text{e}$ and no further energy absorption from the laser $E_{\text{L}_\text{abs}}\equiv 0$,  Eq. \eqref{equ:Energy_conservation_Phi} transforms to Eq. \eqref{equ:Energy_conservation_Phi_Te_const}, which is commonly used in $T_\text{e}$-dependent DFT MD simulations \cite{Silvestrelli1996,Zijlstra2013Jan,Zijlstra2013Aug} at constant $T_\text{e}$.
(2) If the changes of the PES corresponding to $T_\text{e}$ are ignored, {\it i.e.}, $\Phi \equiv \Phi(T_\text{e}=0)$ and $d\Phi / dT_\text{e} =0$, then we recover the commonly used TTM-MD equations
\eqref{equ:e_TTM_MD}, \eqref{equ:ionic_TTM_MD}.
This means that usual TTM-MD aproaches based on empirical interatomic potentials only depending on the atomic coordinates implicitly assume that electrons are always in the ground state.

In order to confirm the validity our method, we performed MD simulations using the $T_\text{e}$-dependent interatomic potential $\Phi^{(\text{Si})}(T_\text{e})$ for Si derived in \cite{Bauerhenne2020} from DFT and compared directly with experimental results by Harb {\it et al.} \cite{Harb2006,Harb2008}.
We solved the ionic equations of motions by using the Velocity Verlet algorithm, which has to be modified \cite{supp-info}.
We used 
\begin{equation}
	G_\text{ep} = 1.8 \times 10^{17} \, \frac{\text{W}}{\text{K}\,\text{m}^3} = 2.19638\times 10^{-8}\, \frac{\text{eV}}{\text{fs}\,\text{K}\,\text{atom}},
	\label{equ:Gep_Si}
\end{equation}
which was also calculated in the framework of DFT \cite{Sadasivam2017}.
Harb {\it et al.} excited free-standing thin Si films by an intense fs laser-pulse with a central wavelength of $\lambda=387$ nm and a FWHM-time width of $\tau=150$ fs \cite{Harb2006,Harb2008}.
In order to compute the energy absorbed by the laser $E_{L_\text{abs}}$, we took the index of refraction $n=6.062+0.630\, i$ of Si at $E_\text{ phot}=3.2$ eV corresponding to $\lambda=387$ nm from the literature \cite{Lide2004} and obtained the absorption coefficient $\alpha_\text{abs}=4\pi\, {\rm Im}(n) / \lambda=0.0204569\, 1/\text{nm}$.
Then we have
\begin{equation}
	E_{L_\text{abs}} = N_\text{at}\left(1 - e^{-\alpha_\text{abs}\, d_\text{film}} \right) \frac{I_{\text{L}_\text{abs}}}{d_\text{film} \, \rho_\text{at}},
	\label{equ:E_L_abs}
\end{equation}
where $d_\text{film}$ is the film thickness, $I_{L_\text{abs}}$ is the absorbed laser fluence at the surface and $\rho_\text{at}=50.8414\, {\rm atoms} / {\rm nm}^3$ is the equilibrium density of Si obtained by us from DFT.

To set up a Si film, we applied periodic boundary conditions in $x$- and $y$-direction and open boundary conditions in $z$-direction ($[\overline{1}11]$ direction of the crystal structure).
We employed the Andersen thermostat \cite{Andersen1980} to initialize the atomic coordinates and velocities at $T_\text{i}=300$ K.
Then we performed MD simulations of the femtosecond-laser excitation (see below) and calculated the time-dependent relative intensities of the Bragg peaks directly from the atomic coordinates \cite{supp-info}.

To study the individual influences of the excited PES with hot electrons and of the EPC on the ionic dynamics, we considered three different scenarios of MD simulations:
(1) excited PES \& EPC: The new method described here (full Eqs. \eqref{equ:Nat_Ce_dTe_dt} and \eqref{equ:ionic_motion_global});
(2) only excited PES: For this we set $G_\text{ep}=0$ in Eqs. \eqref{equ:Nat_Ce_dTe_dt} and \eqref{equ:ionic_motion_global});
(3) only EPC: We used in Eqs. \eqref{equ:Nat_Ce_dTe_dt} and \eqref{equ:ionic_motion_global}) for $\Phi$ the expression $\Phi^{(\text{Si})}(T_\text{e}=0)+E_\text{e}(T_\text{e})$. 
This means that the PES is always evaluated at $T_\text{e}=0$, {\it i.e.}, the bonding corresponds to electrons in their ground state.
The additional term $E_\text{e}(T_\text{e})$ denotes the electronic energy as a function of $T_\text{e}$ derived from DFT for the ideal crystal structure \cite{supp-info}, which is included to obtain the correct $C_\text{e}(T_\text{e})$ from Eq \eqref{equ:C_e_S_e}.

\begin{figure}[htb!]
\includegraphics[width=0.5\textwidth]{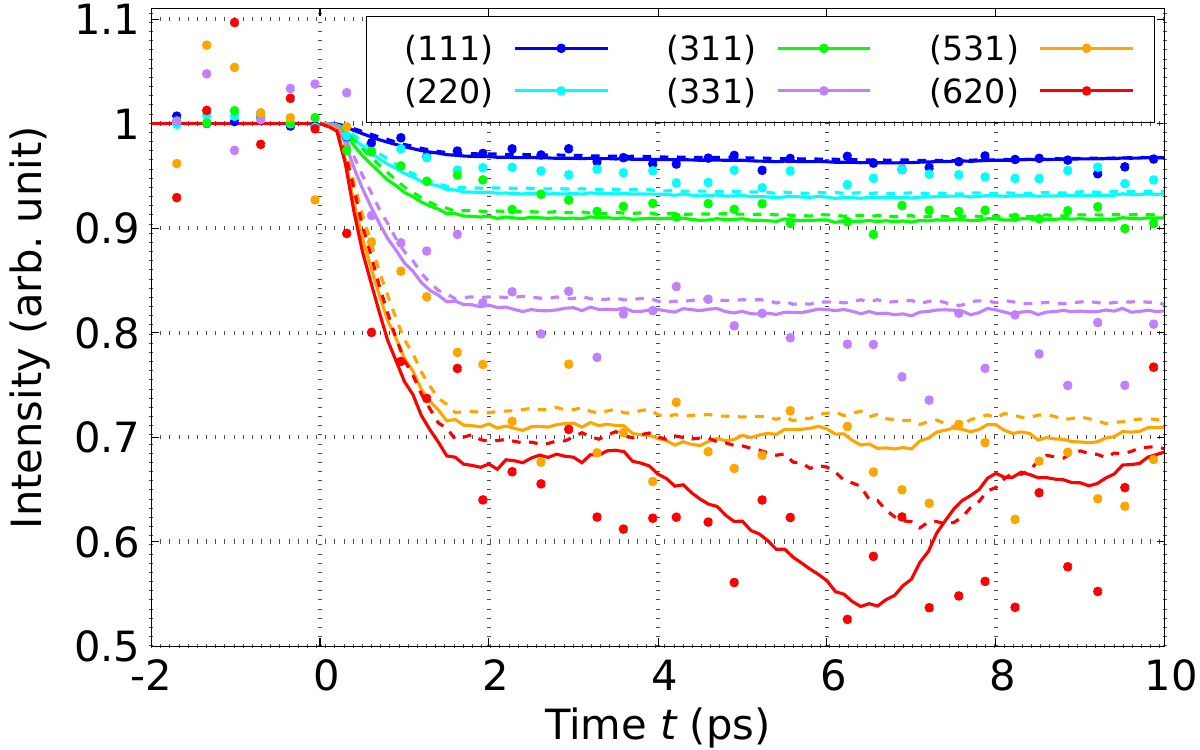}
\caption{Relative intensities of various Bragg peaks are shown as a function of time obtained from the experiment (points) and calculated from our MD simulations considering the excited PES \& EPC (solid lines) and considering only the EPC (dashed lines).	The experimental values are taken from Figure 4 of Ref. \cite{Harb2006}.
\label{fig:Bragg_experiment_100meV}}
\end{figure}

In a first experiment, Harb {\it et al.} excited a Si film with a thickness of $d_\text{film}=50$ nm using a fluence of $I_{\text{L}_\text{tot}}=5.6 \, \text{mJ} / \text{cm}^2$, which is below the damage threshold \cite{Harb2006}.
This corresponds to $E_{L_\text{abs}} / N_\text{at} = 0.1\, {\rm eV} / {\rm atom}$ using Eq. \eqref{equ:E_L_abs}.
By means of ultrafast electron diffraction, they measured the time dependent intensity of various Bragg peaks.
In order to compare directly with this experiment, we set up a simulation cell that consists of $11 \times 11 \times 93$ conventional cells and contains a $50$ nm thick Si film with $N_\text{at}=90024$ atoms.
In Fig. \ref{fig:Bragg_experiment_100meV}, we compare the measured relative intensities with our calculated ones for the six Bragg peaks that Harb {\it et al.} studied.
The relative Bragg peak intensities obtained from the MD simulations considering excited PES \& EPC and considering only the EPC are almost identical and agree with the experimental results.
Thus, here the ionic motion is clearly dominated by the EPC.
Notice: Harb {\it et al.} also derived the time-dependent ionic temperature $T_\text{i}$ of the Si film from the time-dependent Bragg peak intensities using Debye Waller theory.
We also reproduce the measured $T_\text{i}$ \cite{supp-info}.

\begin{figure}[htb!]
	\includegraphics[width=0.5\textwidth]{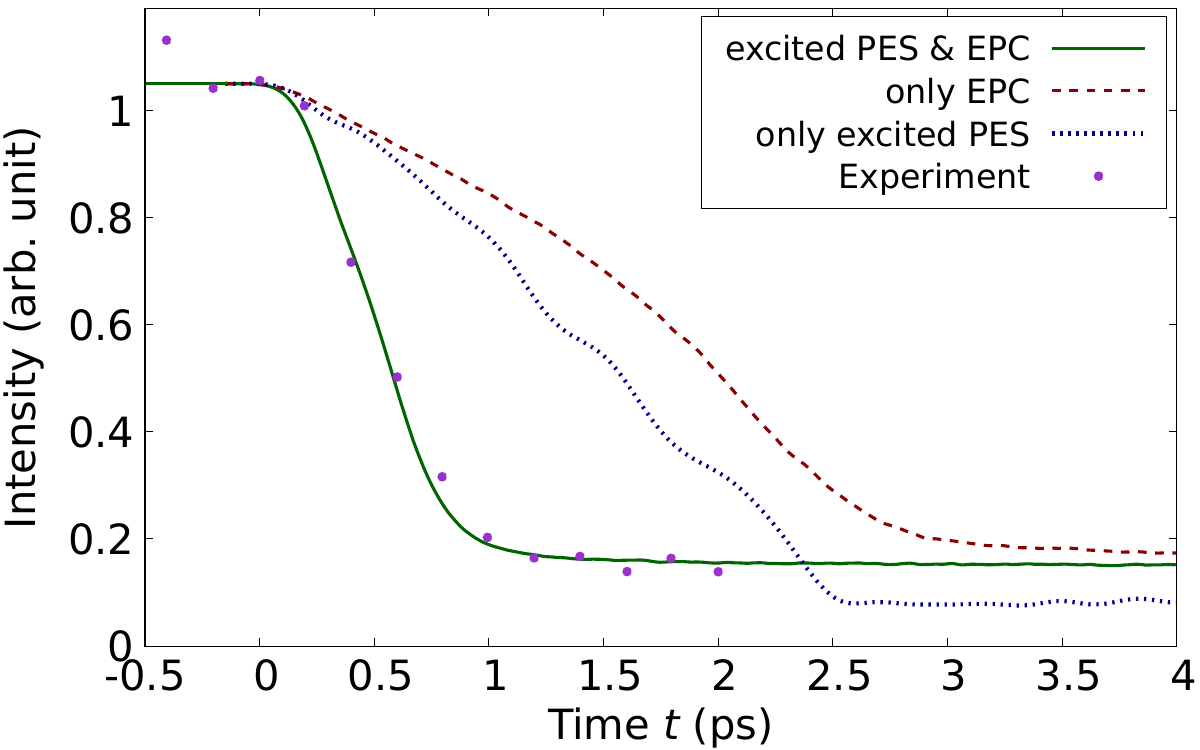}
	\caption{Relative intensity of the (220) Bragg peak is shown as a function of time obtained from the experiment (points) and from our calculations (lines).
	The experimental values are taken from FIG. 3 (c) of Ref. \cite{Harb2008}.
	\label{fig:Bragg_experiment_1200meV}}
\end{figure}

In a second measurement, Harb {\it et al.} excited a Si film with a thickness of $d_\text{film} =30$ nm using a laser fluence of $I_{\text{L}_\text{abs}}=65 \, \text{mJ} / \text{cm}^2$, which is above the damage threshold \cite{Harb2008}.
This corresponds to $E_{L_\text{abs}} / N_\text{at} = 1.2\, {\rm eV} / {\rm atom}$ using Eq. \eqref{equ:E_L_abs}.
We set up a simulation cell that consists of $11\times 11 \times 56$ conventional cells and contains a $30$ nm thick Si film with $N_\text{at}=54208$ atoms.
In Fig. \ref{fig:Bragg_experiment_1200meV} we show the realtive intensity of the (220) Bragg peak obtained from our calculations and from the experiment.
Note, that if only the effect of the excited PES (like in usual DFT approaches) or only the effect of EPC (like in usual TTM-MD simulations using empirical potentials) are assumed, the MD simulations deviate significantly from the experimental results.
If only the excited PES effects are considered, then the Bragg peak intensity shows some oscillations due to the laser-induced movement of the crystal planes against each other and a much slower decay then in experiments.
If, on the other hand, only the EPC effects are considered, the crystal planes melt immediately and no oscillations can be seen.
The decay of the Bragg peak intensities is here also quantitatively and qualitatively different from experiment and is even slower than in the case of considering only the influcence of the excited PES.
In contrast, if the combined effects of the excited PES and the EPC are taken into account using the theory developed in this letter, we obtain a remarkably good agreement with the experimental result.
We point out that no adjusting parameters are used.
The results of Fig. \ref{fig:Bragg_experiment_1200meV} demonstrate that at high fluences both competing contributions, namely the modified PES and the EPC, are essential.

While the theory presented here successfully unifies thermal (incoherent electron-phonon heating) and non-thermal (bond changes) effects in solids post-laser excitation, there remains room for improvement by incorporating electron dynamics during laser excitation and subsequent thermalization.
Such an extension has not been developed so far.

Time dependent density functional theory (TDDFT) \cite{Runge1984,Wijewardane2008,Krishna2009} is indeed able to accurately capture the interaction of the laser pulse with the electrons, but completely fails in describing electron thermalization, limiting its validity roughly to the duration of the laser excitation. In contrast, methods grounded in Boltzmann collision integrals \cite{Kaiser2000, Brouwer2017} successfully model electron thermalization but operate with fixed ionic positions.
We anticipate that the accuracy of the method presented here becomes highly reliable once $T_\text{e}$ has been established,  which typically occurs on a timescale of 50 fs.

In summary, we introduced a unified theory that comprehensively describes laser-induced structural alterations, simultaneously incorporating bond changes and electron-phonon coupling at the microscopic level. Our approach encompasses the TTM-MD and ab-initio DFT simulations as special cases.
When applied to Si, our method demonstrated exceptional agreement with experimental results, affirming its accuracy in capturing the intricate dynamics of laser-induced structural changes.

\begin{acknowledgments}
This work was supported by the DFG through the grant GA 465/18-1 and GA 465/27-1.
Computations were performed on the high-performance computer Lichtenberg  at the NHR Centers NHR4CES at TU Darmstadt, on the IT Servicecenter (ITS) University of Kassel, and on the computing cluster FUCHS University of Frankfurt.
\end{acknowledgments}


%

\end{document}